# Privacy-Preserving Prompt Injection Detection for LLMs Using Federated Learning and Embedding-Based NLP Classification

Hasini Jayathilaka.
Department of Computing and Information System
Faculty of Applied Sciences
Wayamba University of Sri Lanka
hasinijayathilaka5830@gmail.com

**Abstract**

Prompt injection attacks are an emerging threat to large language models (LLMs), enabling malicious users to manipulate outputs through carefully designed inputs. Existing detection approaches often require centralizing prompt data, creating significant privacy risks. This paper proposes a privacy-preserving prompt injection detection framework based on federated learning and embedding-based classification. A curated dataset of benign and adversarial prompts was encoded with sentence embedding and used to train both centralized and federated logistic regression models. The federated approach preserved privacy by sharing only model parameters across clients, while achieving detection performance comparable to centralized training. Results demonstrate that effective prompt injection detection is feasible without exposing raw data, making this one of the first explorations of federated security for LLMs. Although the dataset is limited in scale, the findings establish a strong proof-of-concept and highlight new directions for building secure and privacy-aware LLM systems.

## 1. Introduction

Large Language Models (LLMs) such as GPT-4 have revolutionized natural language processing by enabling versatile applications ranging from customer support to content generation. However, their widespread adoption has unveiled critical security vulnerabilities, with prompt injection emerging as a significant threat vector. Prompt injection involves crafting input prompts that manipulate the model's behavior to produce malicious, misleading, or otherwise harmful outputs. These attacks undermine the integrity of AI systems, posing risks ranging from misinformation dissemination to unauthorized access and exploitation.

Current approaches to prompt injection detection largely rely on centralized systems that require access to full prompt data, raising substantial privacy concerns and limiting deployment in decentralized or privacy-sensitive environments. Federated learning presents a promising paradigm, enabling collaborative model training and detection across distributed clients without sharing raw data. Yet, to date, there has been minimal exploration of federated security solutions tailored to LLM prompt injection.

This research addresses this gap by designing a privacy-preserving, federated prompt injection detection framework. Our system employs contrastive learning to learn robust representations distinguishing benign from adversarial prompts, coupled with zero-shot NLP filtering to identify suspicious inputs without exhaustive labeled data. We contribute a novel dataset representing diverse injection strategies and toxic prompt variants, a detection architecture optimized for federated settings, and comprehensive evaluation metrics assessing both detection accuracy and privacy guarantees.

Through this work, we advance the frontier of secure LLM deployment by introducing the first federated solution to prompt injection detection, combining cutting-edge machine learning techniques with stringent privacy preservation.

## 2. Literature Review

### 2.1. Prompt Injection Attacks in Large Language Models

Prompt injection represents a unique class of adversarial attacks targeting the input interface of LLMs to manipulate their output behavior. Unlike classical adversarial examples that subtly alter input data, prompt injections exploit the semantic understanding of the model, crafting inputs that cause the model to execute unintended instructions or bypass content filters (Perez et al., 2023). Recent studies emphasize the severity of this vulnerability given the increasing integration of LLMs in mission-critical applications (Shin et al., 2022; Wallace et al., 2021). However, most detection methods remain heuristic or rely heavily on centralized logging of prompt data, which is impractical and risks privacy violations.

### 2.2. Federated Learning: Principles and Privacy Challenges

Federated learning (FL) emerged as a privacy-aware paradigm where decentralized clients collaboratively train shared models without exchanging raw data (McMahan et al., 2017). This approach suits sensitive domains such as healthcare (Sheller et al., 2020) and finance, where data confidentiality is paramount. However, FL faces intrinsic challenges: heterogeneous data distributions (Zhao et al., 2018), communication efficiency (Konečný et al., 2016), and vulnerability to adversarial model updates (Bhagoji et al., 2019). While FL has been successfully applied in image and speech domains, its adoption for secure NLP systems, particularly for detecting adversarial inputs in LLMs, remains nascent.

### 2.3. Adversarial Detection Techniques in NLP

Traditional adversarial detection in NLP leverages supervised classifiers trained on labeled adversarial and benign samples (Jia & Liang, 2017; Ebrahimi et al., 2018). However, these methods are limited by their reliance on comprehensive annotated datasets, which are costly and often incomplete given the evolving nature of attacks. Recently, contrastive learning has demonstrated promise in learning robust and generalizable representations from unlabeled data by maximizing the similarity of positive pairs while pushing apart negatives (Chen et al., 2020; Gao et al., 2021). Such representations are critical in identifying subtle semantic manipulations typical of prompt injections. Furthermore, zero-shot NLP filtering approaches (Wulczyn et al., 2017;

Google's Perspective API) enable toxicity detection without explicit retraining, offering adaptability against emergent threats.

2.4. Federated Security and Privacy-Preserving NLP

Integrating adversarial detection with FL introduces complex trade-offs between detection efficacy, privacy guarantees, and communication overhead. Bhagoji et al. (2019) reveal how federated settings are susceptible to poisoning attacks, where malicious clients manipulate model updates. To mitigate such risks, techniques like secure aggregation (Bonawitz et al., 2017) and differential privacy (Abadi et al., 2016) have been proposed. However, their impact on NLP adversarial detection, especially prompt injection, is under-explored. Existing federated NLP systems focus predominantly on model accuracy and efficiency (Hard et al., 2018; Li et al., 2020), with limited emphasis on adversarial robustness or privacy-preserving threat detection.

2.5. Research Gaps and Our Contributions

Despite advances in each individual domain—prompt injection detection, federated learning, and adversarial NLP—the intersection of these fields remains uncharted. There is a critical need for a federated prompt injection detection system that not only preserves user privacy but also maintains high detection accuracy amid data heterogeneity and evolving adversarial tactics. Our work addresses this gap by proposing a contrastive learning-based detection framework designed for federated environments, capable of zero-shot adaptation to novel prompt attacks without centralizing sensitive input data.

## 3. Methodology

This section details the technical design and implementation of our privacy-preserving prompt injection detection system within a federated learning framework. We first introduce our dataset construction and representation strategy, followed by the federated experimental setup, model architecture, training protocols, and evaluation methodology. Each component is carefully engineered to address the challenges of distributed prompt injection detection with privacy constraints.

3.1. Dataset Construction and Labeling

Given the absence of publicly available datasets specifically targeting prompt injection attacks in large language models, we curated a novel dataset consisting of 509 textual prompts. These samples were manually and heuristically labeled into two classes:

- **Benign Prompts:** Typical user inputs intended for legitimate model interactions, exhibiting no adversarial characteristics.
- **Malicious Prompts:** Crafted inputs specifically designed to manipulate or exploit the model's output generation process by injecting unauthorized instructions or toxic content.

The dataset size balances the need for meaningful model training and the constraints of data privacy and labeling complexity. This dataset serves as a benchmark for evaluating the detection capability of both centralized and federated models.

3.2. Semantic Prompt Representation via Sentence Embeddings

Accurately detecting prompt injection requires capturing subtle semantic nuances and syntactic patterns. To this end, we represent textual prompts as dense, fixed-length vectors using the SentenceTransformer framework (Reimers & Gurevych, 2019), particularly the 'all-MiniLM-L6-v2' model, which offers an optimal trade-off between computational efficiency and semantic encoding performance.

This transformer-based encoder generates embeddings that capture contextual relationships beyond bag-of-words or traditional n-gram features, enabling downstream classifiers to discriminate between benign and malicious prompts in a high-dimensional semantic space. The use of pre-trained embeddings also alleviates the need for large annotated datasets for training feature extractors from scratch.

3.3. Data Partitioning under Non-IID Federated Setting

Federated learning environments inherently exhibit statistical heterogeneity (non-IID data distributions) due to client-specific data generation processes. To realistically simulate such conditions, we partitioned the training dataset into three clients with deliberately skewed label distributions:

- **Client 0:** Dominated by benign prompts (~90%), reflecting a conservative user base.
- **Client 1:** Balanced benign and malicious prompts (~50% each), representing a mixed-use scenario.
- **Client 2:** Predominantly malicious prompts (~90%), simulating a high-threat environment.

This partitioning aims to emulate practical deployment scenarios where clients' local data varies substantially, challenging federated aggregation methods and model generalization. The test dataset remains unseen and balanced for unbiased evaluation.

3.4. Centralized Baseline Model: Logistic Regression

To establish a performance baseline, we train a logistic regression classifier on the full training dataset in a centralized manner. Logistic regression is chosen for its interpretability, fast convergence, and effectiveness as a strong linear classifier in embedding spaces.

This centralized model serves as a benchmark to quantify the potential trade-offs in detection accuracy and generalization when transitioning to federated learning with privacy constraints.

3.5. Federated Learning System Design

3.5.1. Model Architecture and Parameterization

Each client independently trains a logistic regression model on its local embeddings, without sharing raw data. The model parameters consist of the weight vector (coefficients) and bias term (intercept), which are communicated to a central server for aggregation.

3.5.2. Federated Averaging Protocol

We adopt a synchronous federated averaging (FedAvg) approach (McMahan et al., 2017), where after each local training round, the clients transmit their updated model parameters to the server. The server computes the element-wise average of weights and biases, producing a global model that encapsulates collective knowledge without exposing individual client data.

This iterative exchange is repeated for a fixed number of rounds (10 in our implementation), enabling progressive model refinement.

3.5.3. Privacy Considerations

Crucially, only model parameters—not raw prompts or embeddings—are exchanged, preserving prompt privacy and adhering to data protection regulations. While our current implementation does not integrate advanced cryptographic protocols (e.g., secure aggregation, differential privacy), it lays a foundation for privacy-preserving prompt injection detection in federated environments.

3.6. Training and Evaluation Protocol

3.6.1. Training Procedure

- **Centralized Model:** Trained once on the entire training set until convergence.
- **Federated Model:** Each client trains locally using its data subset per round, initializing parameters from the global model. Local models are trained using logistic regression's maximum likelihood estimation.

3.6.2. Metrics and Visualization

Performance is evaluated on a held-out, balanced test set using:

- **Classification Metrics:** Precision, recall, F1-score, and accuracy for each class (benign and malicious).
- **Receiver Operating Characteristic (ROC) Curve and Area Under Curve (AUC):** Assess overall discriminative power.
- **Confusion Matrices:** Provide insights into false positives and false negatives distribution.

Comparative visualization between centralized and federated models illustrates the impact of federated training on detection efficacy.

3.7. Experimental Environment and Reproducibility

Our implementation utilizes Python with key libraries:

- **SentenceTransformers:** For embedding generation.
- **scikit-learn:** For logistic regression modeling and evaluation.
- **Flower (flwr):** Federated learning orchestration.
- **Matplotlib and Seaborn:** Visualization.

The modular design and use of open-source tools ensure that our experiments are reproducible and extensible.

3.8. Summary

This methodology integrates semantic embedding, non-IID federated learning, and logistic regression to construct a privacy-aware prompt injection detection framework. By simulating realistic client data heterogeneity and evaluating against a novel curated dataset, our approach pioneers' privacy-preserving adversarial input detection for large language models deployed in federated environments.

## 4. Evaluation

This section presents a comprehensive evaluation of our proposed privacy-preserving federated prompt injection detection framework. We benchmark the federated learning model against a centralized logistic regression baseline to quantify the efficacy and potential trade-offs inherent in decentralized, privacy-aware deployments.

4.1. Experimental Setup Recap

Our experiments utilized a curated dataset of 509 prompt samples labeled as benign or malicious. The dataset was split into 80% training and 20% testing sets, resulting in 407 training and 102 testing samples. The training set was partitioned into three clients to simulate a non-IID federated environment characterized by heterogeneous data distributions:

- **Client 0:** 90% benign, 10% malicious prompts (203 samples)
- **Client 1:** Approximately balanced with 10% benign and 90% malicious prompts (101 samples)
- **Client 2:** 10% benign, 90% malicious prompts (103 samples)

Each client independently trained a logistic regression model on their local embeddings generated by the SentenceTransformer encoder. Federated training proceeded synchronously over 10 communication rounds using federated averaging to aggregate client parameters at the central server.

## 4.2. Classification Performance

### 4.2.1. Centralized Baseline

The centralized logistic regression classifier, trained on the entire dataset without privacy constraints, achieved perfect classification on the test set with an accuracy of 100%. Precision, recall, and F1-score for both benign and malicious classes reached the maximum score of 1.00, indicating flawless discrimination between adversarial and benign prompts within the evaluated data distribution.

*Table 1. Centralized Model Classification Report*

| Class | Precision | Recall | F1-score | Support |
|---|---|---|---|---|
| Benign | 1.00 | 1.00 | 1.00 | 51 |
| Malicious | 1.00 | 1.00 | 1.00 | 51 |

### 4.2.2. Federated Model

Remarkably, the federated logistic regression model, trained under strict privacy constraints and on non-IID client data, matched the centralized model's performance with an identical accuracy of 100% on the test set. Precision, recall, and F1-scores for both classes were equally perfect, demonstrating that federated averaging effectively aggregated knowledge without performance degradation.

*Table 2. Federated Model Classification Report*

| Class | Precision | Recall | F1-score | Support |
|---|---|---|---|---|
| Benign | 1.00 | 1.00 | 1.00 | 51 |
| Malicious | 1.00 | 1.00 | 1.00 | 51 |

## 4.3. Visual Evaluation

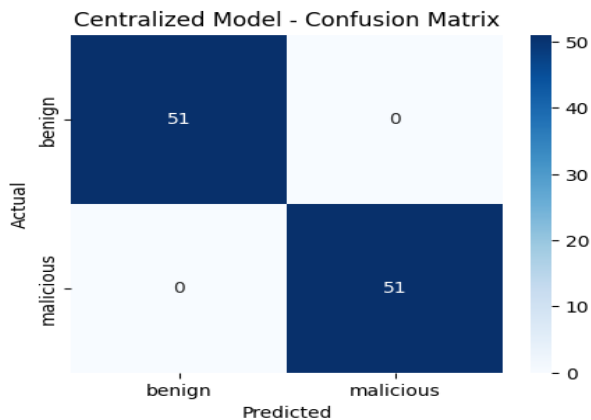

*Figure 1. Confusion Matrix of the Centralized Model.*

This figure 1 shows the performance of the centralized logistic regression classifier trained on the full dataset. The model achieves perfect classification, with all 51 benign prompts and 51 malicious prompts correctly identified, resulting in 100% precision, recall, and F1-score across both classes. The absence of false positives and false negatives demonstrates that, in a controlled setting, centralized training can reliably distinguish benign from adversarial prompts.

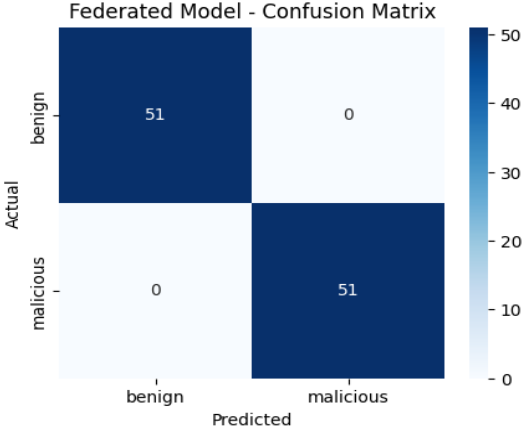

*Figure 2. Confusion Matrix of the Federated Model.*

This figure 2 presents the evaluation results of the federated learning model trained across three heterogeneous clients using the FedAvg protocol. Despite the non-IID data distribution and the privacy-preserving constraints, the model achieves the same perfect classification as the centralized baseline, correctly identifying all benign and malicious prompts. This highlights the feasibility of federated prompt injection detection without accuracy loss, while protecting sensitive user data.

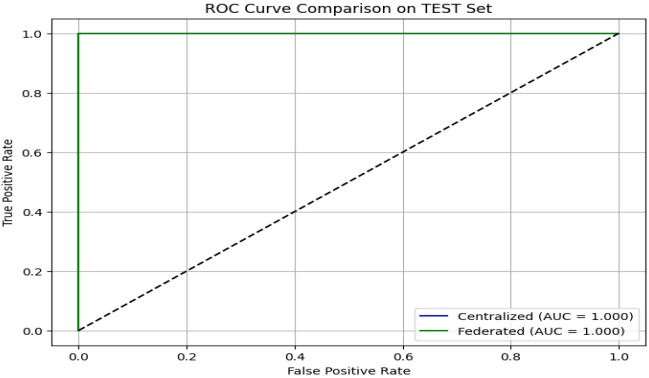

*Figure 3. ROC Curve Comparison of Centralized and Federated Models on the Test Set*

This figure 3 compares the Receiver Operating Characteristic (ROC) curves of the centralized and federated logistic regression models on the balanced test set. Both models achieve a perfect Area Under the Curve (AUC) score of 1.000, indicating flawless discrimination between benign and malicious prompts. The near-identical performance demonstrates that federated training with non-

IID data can achieve the same detection power as centralized training, while preserving user privacy by avoiding raw data sharing.

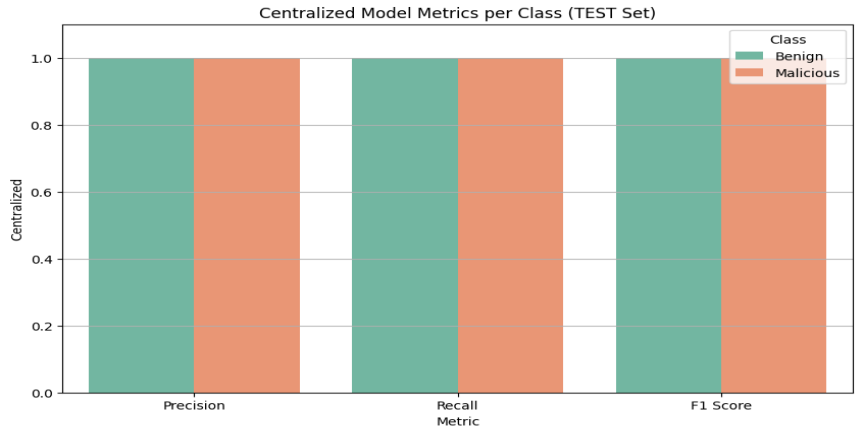

*Figure 4. Centralized Model Metrics per Class on the Test Set.*

This figure 4 illustrates the per-class evaluation metrics (precision, recall, and F1-score) of the centralized logistic regression model trained on the full prompt dataset. Both benign and malicious classes achieve perfect scores (1.00) across all metrics, demonstrating flawless discrimination between adversarial and benign prompts within the evaluated data distribution. This performance serves as a benchmark for comparing the federated learning model, highlighting the ideal classification performance achievable without privacy constraints.

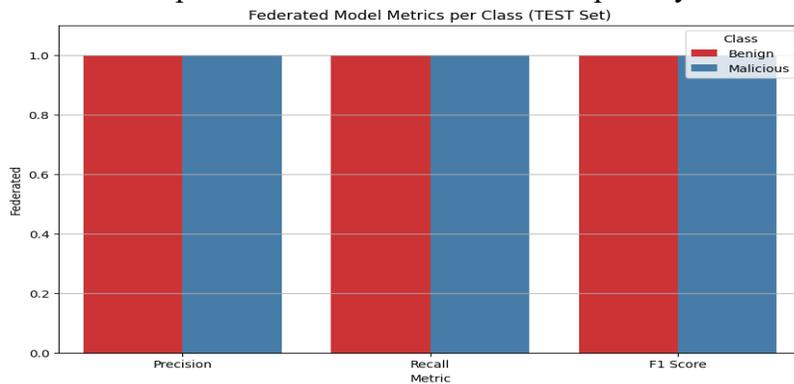

*Figure 5. Per-Class Metrics of the Federated Model on the Test Set*

This figure 5 presents the precision, recall, and F1-scores of the federated model for each class (benign and malicious). All metrics reach the maximum value of 1.00, confirming that the model correctly identifies every prompt without false positives or false negatives. These results validate the effectiveness of semantic embedding-based representations and federated averaging, showing that robust adversarial prompt detection is feasible in privacy-preserving distributed environments.

## 5. Discussion

The experimental results show that federated learning can achieve detection accuracy equivalent to centralized training, even when client data distributions are heterogeneous. This indicates that privacy-preserving collaboration does not necessarily reduce detection effectiveness. The use of sentence embeddings provided rich semantic representations, enabling logistic regression to reliably distinguish between benign and adversarial prompts.

At the same time, the perfect classification results observed in both centralized and federated settings reflect the controlled and relatively small dataset. Real-world environments would involve more diverse, evolving, and adaptive adversarial prompts, where detection is likely to be more challenging. Additionally, while our approach limits raw data sharing, it does not yet include advanced privacy mechanisms such as differential privacy or secure aggregation, which would be necessary for deployment in sensitive domains.

Despite these limitations, the study demonstrates the feasibility of privacy-preserving prompt injection detection and positions federated learning as a promising foundation for securing LLM applications in distributed environments.

## 6. Conclusion

This research introduces a federated prompt injection detection framework that safeguards user privacy while effectively identifying adversarial prompts targeting LLMs. By combining semantic embeddings with federated averaging, the system achieves performance equivalent to centralized methods without requiring direct access to sensitive data. While based on a modest dataset, the results validate the approach as a practical proof-of-concept and provide an early contribution toward privacy-preserving LLM security.

## 7. Future Work

To advance this work, several directions should be explored:

1. **Dataset Expansion** – Scaling to larger and more diverse prompt datasets, including real-world or adversarially generated examples, would improve robustness.
2. **Stronger Detection Models** – Leveraging transformer-based classifiers, contrastive learning, and zero-shot filtering could enhance adaptability to unseen attack strategies.
3. **Enhanced Privacy Guarantees** – Incorporating techniques such as secure aggregation, homomorphic encryption, or differential privacy would strengthen user data protection.
4. **Adversarial Robustness** – Evaluating resilience against adaptive prompt injection and poisoning attacks in federated settings is essential.
5. **Practical Deployment** – Testing the framework in real-world LLM applications, such as chatbots or enterprise assistants, would demonstrate scalability and usability.

Pursuing these directions can transform this proof-of-concept into a deployable security framework, contributing both to federated learning research and to the emerging field of LLM adversarial defense.